\documentclass[
 reprint,
 amsmath,amssymb,
 aps,
showkeys
]{revtex4-2}

\usepackage{graphicx}
\usepackage{dcolumn}
\usepackage{bm}

\usepackage{xspace}
\newcommand{\gao}{Ga\textsubscript{2}O\textsubscript{3}\xspace}
\newcommand{\alo}{Al\textsubscript{2}O\textsubscript{3}\xspace}
\newcommand{\algao}{(Al\textsubscript{x}Ga\textsubscript{1-x})\textsubscript{2}O\textsubscript{3}\xspace}
\newcommand{\algaonox}{(AlGa)\textsubscript{2}O\textsubscript{3}\xspace}
\usepackage{subcaption}
\usepackage{amsmath}
\usepackage{xfrac}
\usepackage{booktabs}

\begin{document}

\preprint{APS/123-QED}

\title{Effect of Annealing on Al Diffusion and its Impact on the Properties of \gao Thin Films Deposited on c-plane Sapphire by RF Sputtering}

\author{Ana Sofia Sousa}
\email{ana.sofia.sousa@tecnico.ulisboa.pt}
\author{Duarte M. Esteves}
\affiliation{Instituto de Engenharia de Sistemas e Computadores --- Microssistemas e Nanotecnologias (INESC MN), Portugal;\\
Instituto de Plasmas e Fusão Nuclear, Instituto Superior Técnico, Universidade de Lisboa, Portugal.}

\author{Tiago T. Robalo}
\author{Mário S. Rodrigues}
\affiliation{Departamento de Física, Faculdade de Ciências, Universidade de Lisboa, Portugal;\\
\mbox{BioISI—Biosystems and Integrative Sciences Institute, Faculdade de Ciências, Universidade de Lisboa, Portugal.}}

\author{Luís F. Santos}
\affiliation{Centro de Química Estrutural, Institute of Molecular Sciences and Departamento de Engenharia Química, Instituto Superior Técnico, Universidade de Lisboa, Portugal.}

\author{Katharina Lorenz}
\author{Marco Peres}
\email{marcoperes@ctn.tecnico.ulisboa.pt}
\affiliation{Instituto de Engenharia de Sistemas e Computadores --- Microssistemas e Nanotecnologias (INESC MN), Portugal;\\
Instituto de Plasmas e Fusão Nuclear, Instituto Superior Técnico, Universidade de Lisboa, Portugal;\\
\mbox{Departamento de Engenharia e Ciências Nucleares, Instituto Superior Técnico, Universidade de Lisboa, Portugal.}}

\begin{abstract}

Gallium oxide is a wide-bandgap semiconductor which has been steadily growing in popularity due to its ultra-wide bandgap, suitability for harsh environments and distinctive opto-electrical properties. Notable applications include deep-UV photodetectors, low loss waveguides or even transparent solar cells. RF sputtering stands out among possible techniques for the epitaxial deposition of \gao thin films with high quality and at a low cost. By using sapphire substrates, and through thermal annealing, we can form a $\beta$-\algao alloy by Al diffusion, which has tunable opto-electrical properties such as the bandgap and breakdown electric field.

In this work, techniques such as X-ray diffraction, Rutherford backscattering spectrometry, Raman spectroscopy, atomic force microscopy and optical transmission are used to determine the optical properties, morphology and composition of \gao deposited and annealed thin films. To explore the formation of the $\beta$-\algao alloy, annealing was performed at variable temperature, in ambient air. It was determined that the bandgap can indeed be tuned between 4.85 and 5.30 eV by varying the annealing temperature, corresponding to an Al content between 0--68.5\%.

\end{abstract}

\keywords{\gao, thin films, RF sputtering, interdiffusion, $\beta$-\algaonox}

\maketitle

\section{Introduction}

Gallium oxide is an emerging Wide-Bandgap Semiconductor (WBS) which has gained a lot of attention in recent years. As other WBSs, it outperforms traditional semiconductors in high-power electronics and high-transparency applications. In particular, \gao stands out for its combination of exceptional optical and electrical properties, such as high breakdown electric field of 8 MV/cm, ultra-wide bandgap of 4.85 eV at room temperature \cite{pearton_review_2018} and high electrical conductivity \cite{orita_preparation_2002} which can be tuned through doping \cite{higashiwaki_-ga2o3_2022}. In this context, it is worth noting that \gao is one of the few semiconductors that can exhibit high transparency in the ultraviolet region along with high electrical conductivity \cite{pearton_review_2018}. In addition to these unique characteristics, \gao exhibits high thermal and chemical stability \cite{kaur_strategic_2021}. Like most wide bandgap semiconductors, it also demonstrates low susceptibility to displacement damage caused by particle irradiation \cite{wong_radiation_2018, higashiwaki_-ga2o3_2022}. These properties make \gao a promising material for the development of devices designed for harsh environment applications. When compared with other competitive semiconductors, such as silicon carbide (SiC) or gallium nitride (GaN) crystals \cite{higashiwaki_gallium_2020, pearton_review_2018}, high-quality thin films of \gao can be easily deposited epitaxially \cite{jessen_supercharged_2021} and high-quality single crystals can be grown by cost-effective melt-growth processes.

Its main applications include schottky barrier diodes, field effect transistors and deep-ultraviolet (DUV) photodetectors \cite{liu_review_2019, xu_gallium_2019, higashiwaki_gallium_2020}. In particular, \gao is an excellent candidate for a new generation of DUV photodetectors due to its solar blind response and shorter absorption cut-off edge \cite{xu_gallium_2019, kaur_strategic_2021}, and because it is possible to tune its bandgap by alloying it with aluminum \cite{liao_wide_2021}. Other remarkable uses for this material include the photo-electrolysis of water, gas sensors, magnetic or resistance random access memories, luminescent displays, waveguides \cite{zhou_demonstration_2019} or even in the development of transparent solar cells \cite{ramana_properties_2019, saikumar_reviewrf_2019, pearton_review_2018, schurig_optimizing_2019}.

In this work, we are focused on the potential of \gao thin films deposited by radio-frequency (RF) magnetron sputtering, a very popular physical vapor deposition technique. This main advantages of this method are its speed, simple equipment requirements and high-quality coating over large areas, on different types of substrates \cite{xu_gallium_2019}.

Despite its advantages as a physical deposition technique,  previous studies have shown that deposition at room temperature using RF sputtering typically results in \gao amorphous films, independently of the crystalline structure of the substrate or gas conditions \cite{ramana_properties_2019}. It is consensual that post-thermal treatments or deposition at higher temperatures are required to promote the crystallization of \gao thin films deposited by RF sputtering \cite{ramana_properties_2019, mishra_effect_2022}. While various studies have already demonstrated the crystallization promoted by post-thermal treatments, few have addressed the interdiffusion processes driven by annealing between the deposited film and the substrate. This process is particularly relevant for \gao films deposited on sapphire, not only because previous studies have demonstrated gallium diffusion into the sapphire substrate \cite{apostolopoulos_diffusion_2006}, but also because the diffusion of aluminum into the \gao films cannot be ruled out. Such diffusion may significantly alter the optical properties of the films, particularly by increasing the bandgap of \gao. In this context, this study aims to investigate and interpret the compositional, structural, morphological and optical changes that occur during annealing, taking into account the interdiffusion processes of Al and Ga between the \gao film and the \alo substrate.

\section{Results \& Discussion}

The \gao thin film was deposited at room temperature, on $c$-plane sapphire substrate wafer. Its thickness was determined to be $(118 \pm 3)$ nm using a surface step profile analyzer. The wafer was then cut, in order to create a set of seven samples. These were then annealed in air ambient, at 550--1300 °C, for 1 hour, using a tubular furnace.

Figure~\ref{fig:rbs-temperature-profile} shows the Rutherford backscattering spectrometry (RBS) spectra obtained for all of the pieces. Through fitting the spectrum of the as-deposited sample considering the depth profile distribution presented in Figure~\ref{fig:rbs-temperature-depth}, the Ga/O/Fe ratio of this as-deposited thin film was determined to be \mbox{38.8\%} of gallium, 58.2\% of oxygen and 2.9\% of iron. From previous studies, we know that this iron contamination stems from the sputtering of material from the magnetron shield, which is made of stainless steel. For the fitting of the annealed samples, this composition was kept fixed, and multiple layers were introduced (see Figure~\ref{fig:rbs-temperature-depth}) to simulate interdiffusion between the \gao film and the \alo substrate. This analysis clearly demonstrates that the aluminum composition increases with the annealing temperature, starting from the onset of diffusion at $\sim$ 850 °C and reaching a maximum $x$ of 68.5\% at 1300 °C. Simultaneously, a clear diffusion of Ga into the \alo substrate is also observed. This experimental evidence of the interdiffusion process, observed in this work for the first time by RBS, is consistent both with previous studies on the diffusion of Ga into the \alo substrate \cite{apostolopoulos_diffusion_2006} and with more recent studies on the diffusion of aluminum from the \alo substrate into \gao films \cite{li_impact_2020, liao_wide_2021}. From this analysis, in addition to the interdiffusion process, there is also strong evidence that the annealing temperature promotes a significant increase of the roughness, as reflected in the estimated values (see Table~\ref{tab:summary}).

Figure~\ref{fig:afm-scans} shows the atomic force microscopy (AFM) images (5 \textmu m $\times$ 5 \textmu m scan) of the as-deposited and annealed samples at 550, 1000 and 1300 °C, where it can be seen that the films present a smooth surface for the as-grown sample. The root mean square roughness estimated from these images increases markedly with the annealing temperature --- from 0.5 to 8 nm after annealing at \mbox{1300 °C}. This increase in roughness may be associated with the growth and coalescence of \gao grains promoted by the annealing temperature \cite{guo_growth_2014, dong_effects_2016}. Considering the RBS results, the effect of aluminum incorporation cannot be ruled out. Indeed, and in accordance with previous studies in \algao films with different aluminum concentration, a significant increase in roughness was observed --- approximately sixfold when the aluminum concentration $x$ increases from 0 to 72\% \cite{wang_temperature_2016}.

Figure~\ref{fig:xrd-temperature-big} shows the X-ray diffraction (XRD) $2\theta$-$\omega$ curves taken of the sample as-deposited and of the samples annealed at different temperatures. The thin films appear to be highly $\Bar{2}01$ textured when starting to crystallize at \mbox{$\sim$ 700 °C}, as practically only the $\Bar{2}01$, $\Bar{4}02$ and $\Bar{6}03$ peaks are visible. These peaks become more intense and narrower as the temperature increases, demonstrating increased crystalline quality and/or larger grain size. However, for the samples annealed at 1300 °C, the texture changes, and more peaks become visible. In particular, a significant increase in the intensity of the peaks associated with the $\{100\}$ family of planes is observed.

Figure~\ref{fig:xrd-temperature-zoom} provides a more detailed view of the effect of annealing temperature on the shift of the diffraction peaks, particularly the $\Bar{6}03$ peak. From the position of these peaks, the interplanar spacing $d$ can be determined through \cite{liao_wide_2021}

\begin{equation}
  \frac{1}{d^2} = \frac{h^2}{a^2 \sin^2 \beta} + \frac{l^2}{c^2 \sin^2 \beta} - \frac{2hl\cos \beta}{ac\sin^2 \beta} + \frac{k^2}{b^2},
\label{eq:d-gao}
\end{equation}

\noindent and for (Al\textsubscript{x}Ga\textsubscript{1-x})\textsubscript{2}O\textsubscript{3} \cite{liao_wide_2021},

\begin{equation}
  \begin{cases}
    a &= (12.21 - 0.42x)\; \text{\AA}\\
    b &= (3.04 - 0.13x)\; \text{\AA}\\
    c &= (5.81 - 0.17x)\; \text{\AA}\\
    \beta &= (103.87 + 0.31x)^\circ
\end{cases},
\label{eq:params-gao}
\end{equation}

\noindent from which the aluminum content $x$, can be estimated. The $h$, $k$ and $l$ correspond to the Miller indices, and $a$, $b$, $c$ and $\beta$ correspond to the \gao lattice parameters. \newpage \noindent The choice of the $\Bar{6}03$ peak is justified by providing the best angular resolution out of the measured peaks, and the fact that it has no overlap with any peaks from the substrate.

This data can also be used to ascertain how the crystal domain size $\tau$ and microstrain $\varepsilon$ of the films are affected by annealing, by analyzing the width of the peak. The crystal domain size can be estimated by using the Scherrer equation \cite{scherrer_bestimmung_1918, harrington_back--basics_2021},

\begin{equation}
  \beta_\tau = \frac{K \lambda}{\tau \cos \theta_{hkl}},
\label{eq:scherrer}
\end{equation}

\noindent where $\lambda$ is the X-ray wavelength, $\beta_\tau$ is the full width at half maximum (FWHM) (after correcting for the instrumental broadening, in radians), $\theta_{hkl}$ is the Bragg angle of the $hkl$ reflection and $K$, the Scherrer constant, is a shape factor usually taken to be $0.9$. The microstrain within the grains is given by the Wilson equation \cite{stokes_diffraction_1944, harrington_back--basics_2021},

\begin{equation}
  \beta_\varepsilon = 4 \varepsilon \tan \theta_{hkl},
\label{eq:wilson}
\end{equation}

\noindent where $\beta_\varepsilon$ is the FWHM of the peak, (corrected for the instrumental broadening, in radians). The peaks were fitted with a pseudo-Voigt function, with Lorentzian parameter $\eta$. 

Assuming that the Gaussian and Lorentzian components of the broadened profile are due, respectively, to grain size broadening $\beta_\tau$ and microstrain broadening $\beta_\varepsilon$ \cite{de_keijser_determination_1983, metzger_defect_1998}, and by using the approximations determined by De Keijser \textit{et al.} \cite{de_keijser_determination_1983}, it follows that

\begin{widetext}
\begin{equation}
    \tau = \frac{K \lambda}{\beta \left(0.017475 + 1.500484 \eta - 0.534156 \eta^2\right) \cos \theta_{hkl}},
    \label{eq:tau}
\end{equation}

\begin{equation}
    \varepsilon = \frac{\beta \left(0.184446 + 0.812692 (1-0.998497 \eta)^{\frac{1}{2}}-0.659603 \eta + 0.445542 \eta^2\right)}{4 \tan \theta_{hkl}},
    \label{eq:eps}
\end{equation}
\end{widetext}

\noindent where $\beta$ is simply the integral breadth of the peak, easily determined from the fit.

Figure~\ref{fig:xrd-pv} shows how the crystal domain size $\tau$ tends to increase with the annealing temperature, which indicates that the grains within the film are increasing in size. The microstrain $\varepsilon$, which is caused by defects and/or inhomogeneity within these grains, is decreasing as the annealing temperature increases. Overall, this analysis demonstrates that there is an increase in the crystalline quality and grain size of the thin film with annealing temperature.

To complement the structural analysis by XRD, a characterization by Raman spectroscopy was then performed. Figure~\ref{fig:raman} shows clear evidence of the A$_\text{g}^\text{(3)}$ and A$_\text{g}^\text{(10)}$ peaks \cite{kranert_lattice_2015} expected for pure $\beta$-\gao, in the samples annealed at temperatures higher than $\sim$ 700 °C,as well as A$_\text{1g}$, E$_\text{g}$ and other peaks belonging to the sapphire substrate \cite{aminzadeh_raman_1999}. Beyond the identification of these peaks, as well as their intensity increase with the annealing temperature, which clearly suggests an improvement in crystalline quality, their shift to higher wavenumbers was also observed. According to Kranert \textit{et al.} \cite{kranert_lattice_2015}, this shift and broadening of the A$_\text{g}^\text{(3)}$ peak is due to increasing aluminum concentration in the \algao films. While this result confirms the aluminum diffusion already shown by the RBS and XRD analyses, it is equally important to emphasize that, even for the sample annealed at \mbox{1300 °C}, with a high concentration of aluminum the A$_\text{g}^\text{(3)}$ peak remains influenced by contributions at lower wavenumbers expected for pure $\beta$-\gao without aluminum. This fact strongly suggests that the Raman signal originates from regions with varying aluminum concentration, supporting the depth heterogeneity estimated by the RBS analysis.

Finally, optical transmission (OT) measurements were conducted, as shown in Figure~\ref{fig:temperature-tauc}, in order to estimate the optical bandgap of the samples. This was done using Tauc's method, wherein \cite{wood_weak_1972}

\begin{equation}
    \alpha \cdot h \nu = (h \nu - E_\text{g})^{\sfrac{1}{2}}
\end{equation}

\noindent for a direct bandgap semiconductor, where $h\nu$ is the energy of the incident photon, $\alpha$ the absorption coefficient and $E_\text{g}$ the bandgap energy. Although \gao technically has an indirect bandgap, at \mbox{$E_\textbf{g} = 4.66\; \text{eV}$}, the direct transition is so close at $4.69\; \text{eV}$ \cite{pearton_review_2018} that we can consider it, at room temperature, to be a direct bandgap.

In Figure~\ref{fig:41-bandgap}, the dependence of the estimated bandgap energy values on the annealing temperature is presented up to 1150 °C, due to setup constraints. In agreement with the work of Li \textit{et al.} \cite{li_impact_2020} the clear trend of increasing bandgap energy from $\sim$ 4.9 to 5.3 eV with the annealing temperature, is, once again, a strong evidence of aluminum diffusion from the substrate into the \gao film. Considering the possible effects of quantum confinement, as already observed in other oxide semiconductors such as ZrO\textsubscript{2}, TiO\textsubscript{2}, ZnO, and WO\textsubscript{3} \cite{gullapalli_structural_2010, ramana_size-effects_2009, tan_blueshift_2005, may_optical_2007}, the annealing would be expected to reduce the bandgap, given the improvement in crystalline quality and increase in grain size; this further supports the aluminum diffusion observed.

Figure~\ref{fig:transmission} demonstrates an increase in the bandgap with both the annealing temperature and annealing time, placing the threshold for diffusion at \mbox{$\sim$ 700 °C}, in agreement with the results discussed before.

Based on the bandgap energy estimated using the Tauc method, the Al content was also estimated by applying Vegard's law, as defined by equation \cite{liao_wide_2021},

\begin{equation}
\begin{aligned}
  E_\text{g}(x) =& (1-x) E_\text{g} (\beta\text{-Ga}_\text{2}\text{O}_\text{3}) + \\
  &+ x E_\text{g} (\theta\text{-Al}_\text{2}\text{O}_\text{3}) - bx(1-x),
\end{aligned}
\label{eq:diffusion-bandgap}
\end{equation}

\noindent where $E_\text{g} (\beta\text{-Ga}_\text{2}\text{O}_\text{3})$ and $E_\text{g} (\theta\text{-Al}_\text{2}\text{O}_\text{3})$ are the bandgaps for monoclinic $\beta$-\gao and $\theta$-\alo, ($4.88$ and $7.24$ eV, respectively), $b =  0.93\; \text{eV}$ is the bowing parameter and $x$ is the Al content \cite{liao_wide_2021}. Solving the equation numerically then yields $x$.

Figure~\ref{fig:comparison} shows the concentration of aluminum of the samples obtained through the different techniques: RBS, XRD and OT. These agree well and show the common trend of Al/Ga interdiffusion starting at \mbox{$\sim$ 700 °C}. Small discrepancies are attributed to the compositional gradients with depth, which influence the results of each technique in different ways.

\section{Conclusions}

In conclusion, from a complementary analysis using different characterization techniques --- RBS, XRD and Raman spectroscopy --- it was clearly demonstrated that as the annealing temperature increases, the aluminum content in the \gao thin films deposited by RF sputtering on \alo substrates also increases. It was shown that the incorporation of aluminum is due to an interdiffusion effect, with aluminum from the substrate moving into the film and gallium from the film moving into the substrate. In addition to the significant increase in roughness with the annealing temperature, a significant improvement on the crystalline structure associated with a preferential orientation grain growth was also observed with the annealing temperature. However, it should be highlighted that for higher annealing temperatures (\mbox{$\sim$ 1300 °C}) the $\Bar{2}01$ preferential orientation was observed to be lost. Beyond the effects on morphological and structural properties, the diffusion of aluminum into the film during thermal annealing, despite not being homogeneous in depth, has a significant impact on the optical properties of the deposited films. OT measurements revealed a higher shift of approximately 0.4 eV for the sample annealed at \mbox{1150 °C}. This study is particularly relevant not only because sapphire is one of the most commonly used substrates for thin film deposition, but also because, in the case of the $\beta$-\gao semiconductor, the incorporation of aluminum may play a key role in the development of applications in the deep ultraviolet region.

\section{Experimental Section}

The \gao thin film was deposited by RF sputtering, performed in a home-built sputtering chamber. These depositions were carried out at 60 W, 6 mTorr and at room temperature, with the sample plate placed approximately 7.3 cm above the 2$''$ \gao target (99.99\% purity), with an argon flow of 15 sccm, for 2 hours. A quarter of $c$-plane  4$''$ sapphire wafer was used, which had been previously cleaned with acetone, isopropyl alcohol and deionized water.

Profilometry measurements were then performed, using a Dektak XT, with a 2.5 \textmu m stylus, with a scan speed of 6.67 \textmu m/s. These were done on 9 points, distributed on a 3 by 3 grid throughout the sample, so as to determine its average thickness.

The thin films samples were cut into into smaller pieces, using a diamond-tipped pen. A study was then performed, by annealing at a variable temperature, from 550 to 1300 °C, at 150 °C steps, for 1 hour. These annealings were performed in air atmosphere, using a tubular furnace.

Rutherford backscattering spectrometry was then performed, using the 2.5 MV Van de Graaff accelerator at Instituto Superior Técnico \cite{alves_insider_2021} to generate the 2 MeV He\textsuperscript{+} beam used to probe the samples. A Si pin diode detector was used (nominal resolution 15 keV), placed at 165° with respect to the beam incident direction, and the samples were tilted by 60° towards it in order to enhance the depth resolution. The data were fitted using WiNDF \cite{barradas_advanced_2008}, to ascertain the depth composition profile of the samples, and considering their surface roughness.

X-ray diffraction followed, in order to probe the crystalline structure of the thin films. These measurements were acquired on a Bruker D8 Discover Diffractometer, with a copper target and tungsten anode, operated at \mbox{40 kV} \mbox{40 mA}, in low resolution mode. The primary beam was parallelized using a parabolic Göbel mirror, and it was collimated using a 0.6 mm slit; the diffracted beam was collimated using a Soller slit and measured using a scintillation detector. These samples were aligned according to the main substrate peak \cite{harrington_back--basics_2021}. The instrumental broadening was estimated to be that of the $K_{\alpha1}$ of the sapphire substrate. In order to determine the precise position of the $\Bar{6}03$ film peaks, these measurements were repeated and corrected for any deviations according to the $006$ sapphire peak.

Atomic force microscopy measurements were then realized, on a PicoLE Molecular Imaging AFM, using a cantilever with a spring constant of 5.4 N/m, to avoid damaging the sample. The measurements were performed in tapping mode, analyzing a 5$\times$5 \textmu m area at a time. The images were then processed using Gwyddion \cite{necas_gwyddion_2012}.

Raman spectroscopy measurements were done using a confocal LabRAM HR 800 Evolution micro-Raman spectrometer, to gain some insight into the structure of the film. Data was acquired in the range of 200--1800 nm, using an external diode laser with a wavelength of \mbox{532 nm} to excite the sample, a 100$\times$ objective and $\sim$ 10 mW laser power. 

Finally, optical transmission measurements were performed to determine the samples' optical bandgap. These were done on an SE-2000 Spectroscopic Ellipsometer, with a spectral range of 190 to 2100 nm. The substrate was used as a reference, and the measurements were subsequently normalized.

\newpage

\section*{Acknowledgements}

The authors acknowledge the financial support from the Portuguese Foundation for Science and Technology (FCT) via the IonProGO (2022.05329.PTDC, http://doi.org/10.54499/2022.05329.PTDC) project, Research Unit INESC MN (UID/05367/2020) acknowledges FCT funding through pluriannual BASE and PROGRAMATICO. D. M. Esteves thanks FCT for his PhD grant (2022.09585.BD).

\begin{widetext}

\begin{figure}[h]
    \centering
    \subfloat[\label{fig:rbs-temperature-profile}]{\includegraphics[width=0.48\linewidth]{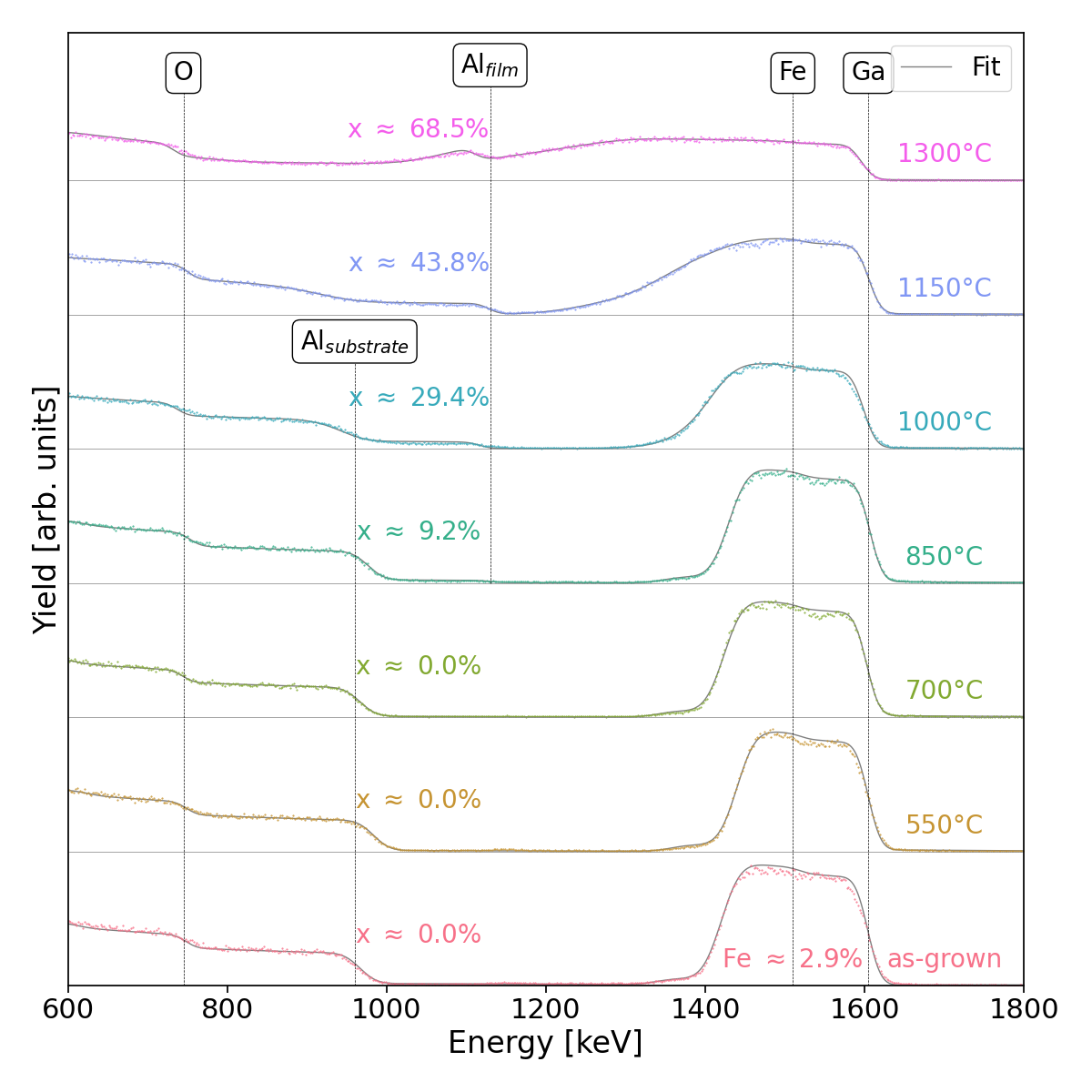}}\hfill
    \subfloat[\label{fig:rbs-temperature-depth}]{
    \includegraphics[width=0.48\linewidth]{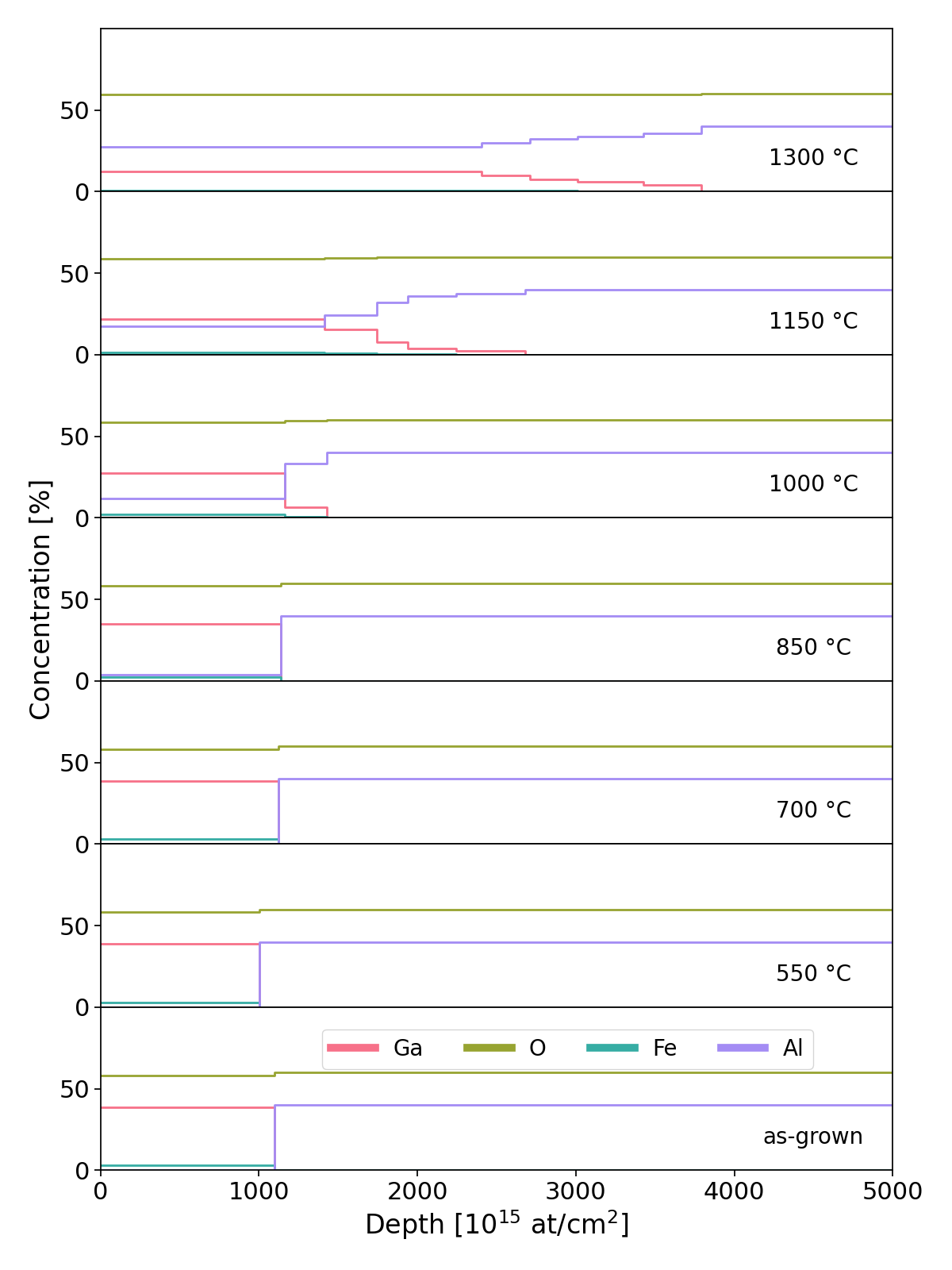}}
    \caption{\textbf{(a)} RBS spectra and respective fits, and \textbf{(b)} depth distributions of the elements considered for the fits of as-grown and annealed samples at 550--1300 °C, for 1 hour. In \textbf{(a)}, the channels corresponding to O, Al (Al\textsubscript{film}), Fe and Ga are marked as well as the barrier as corresponding to the substrate before diffusion.}
    \label{fig:rbs-temperature}
\end{figure}

\begin{figure}[h]
  \centering
  \subfloat[\label{fig:afm-asgrown}] {\includegraphics[width=0.32\linewidth]{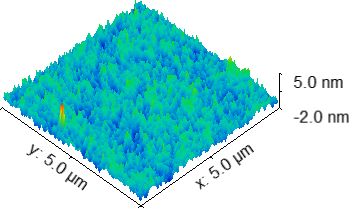}}\hfill
  \subfloat[\label{fig:afm-550}] {\includegraphics[width=0.32\linewidth]{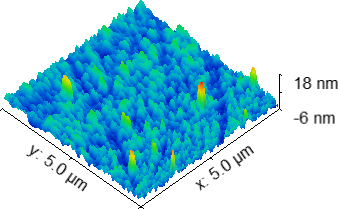}}\\
  \subfloat[\label{fig:afm-1000}] {\includegraphics[width=0.32\linewidth]{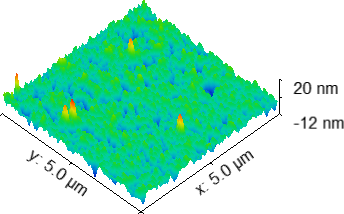}}\hfill
  \subfloat[\label{fig:afm-1300}] {\includegraphics[width=0.32\linewidth]{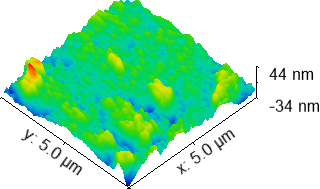}}
  \caption{AFM images of the \textbf{(a)} as-deposited sample \textbf{(b)} annealed at 550 °C , \textbf{(c)} 1000 °C  and \textbf{(d)} 1300 °C.} \label{fig:afm-scans}
\end{figure}

\begin{table*}[h]
\centering
\caption{Comparison of the Al content $x$ obtained through the different techniques and roughness, for the different annealing temperatures.}
\label{tab:summary}
\begin{tabular}{ccccc}
\toprule
Annealing Temperature [°C] & $x$ XRD [\%] & $x$ RBS [\%] & $x$ Tauc [\%] & AFM RMS roughness [nm] \\
\midrule
-- & -- & $0.0$ & $-2 \pm 4$ & $0.5$ \\
550 & -- & $0.0$ & $-2 \pm 5$ & $2.1$ \\
700 & $0.7 \pm 0.9$ & $0.0$ & $1 \pm 5$ & -- \\
850 & $7.3 \pm 0.4$ & $9.2$ & $6 \pm 4$ & -- \\
1000 & $23.4 \pm 0.2$ & $29.4$ & $19 \pm 6$ & $2.7$ \\
1150 & $42.02 \pm 0.07$ & $43.8$ & $24 \pm 9$ & -- \\
1300 & $66.80 \pm 0.07$ & $68.5$ & -- & $8.3$ \\
\bottomrule
\end{tabular}
\end{table*}

\begin{figure}[htb]
\centering
  \subfloat[\label{fig:xrd-temperature-big}]{\includegraphics[width=0.5\linewidth]{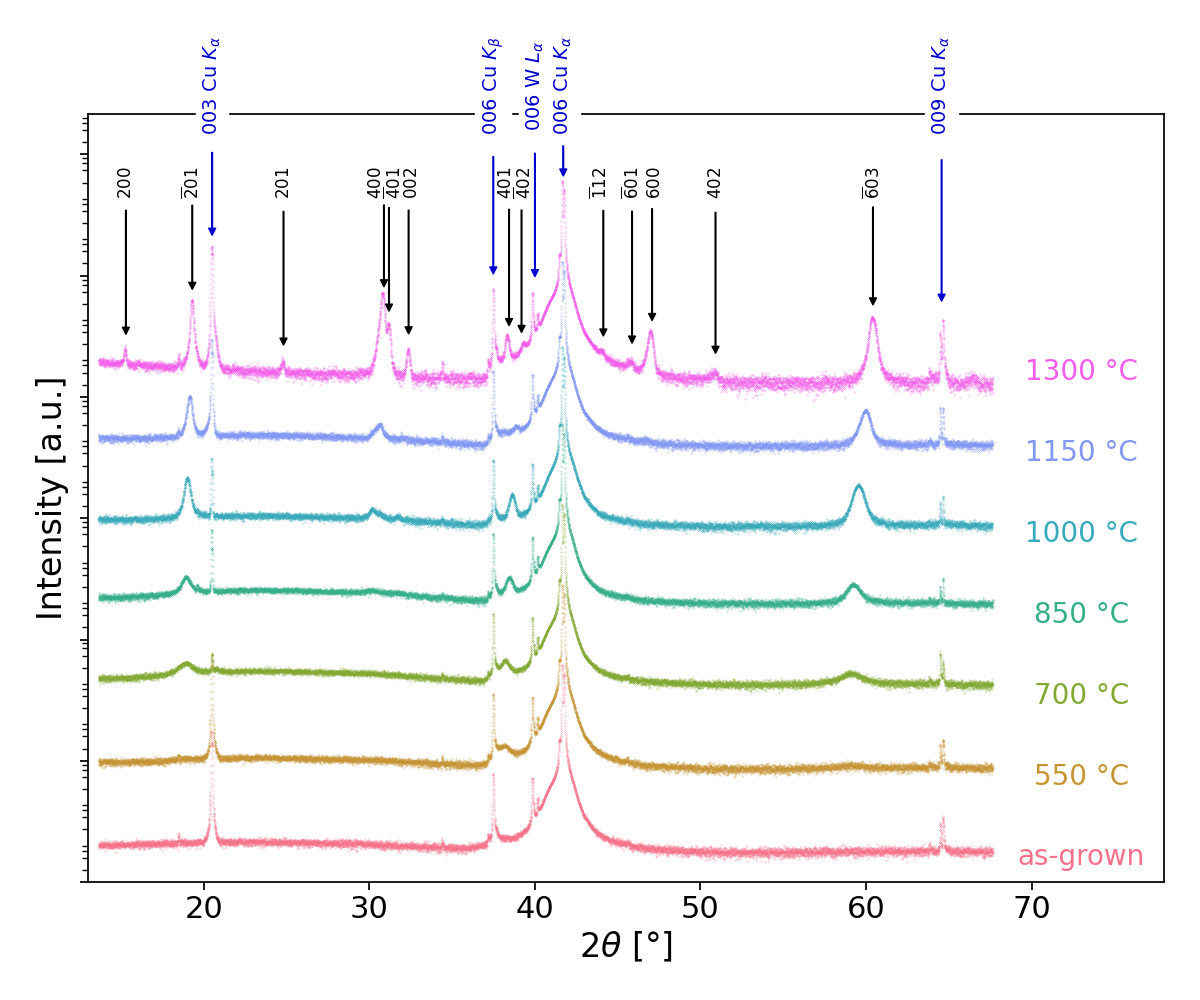}}
  \subfloat[\label{fig:xrd-temperature-zoom}]{\includegraphics[width=0.32\linewidth]{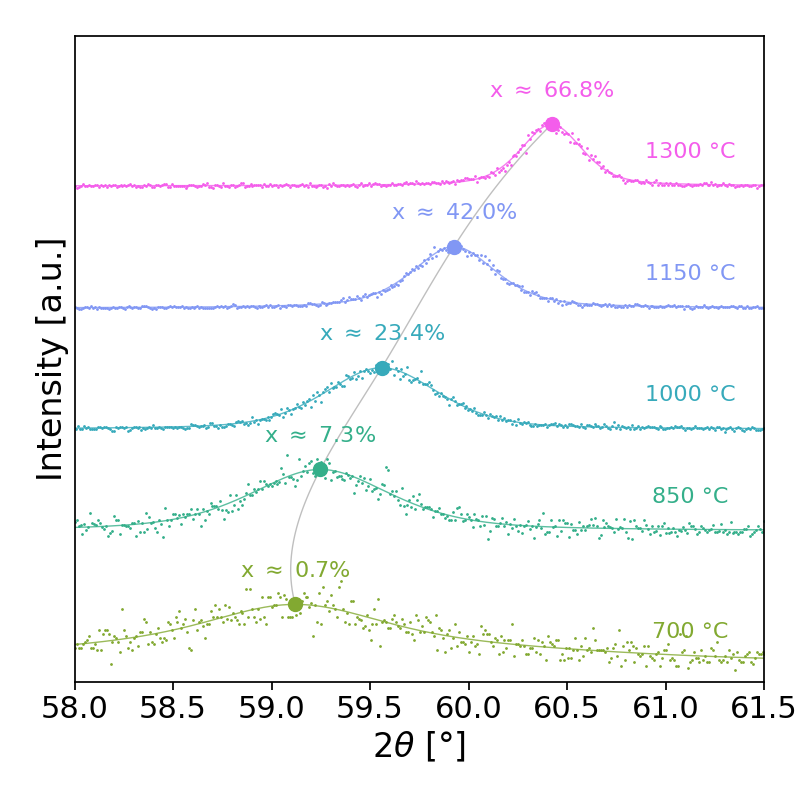}}
  \caption{XRD $2\theta$-$\omega$ curves, as-grown and annealed at 550 -- 1300 °C, for 1 hour \textbf{(a)} of all peaks in the range 14--68° and \textbf{(b)} showing in detail the $\Bar{6}03$ peak.}
  \label{fig:xrd-temperature}
\end{figure}

\begin{figure}[h]
  \centering
  \includegraphics[width=0.4\textwidth]{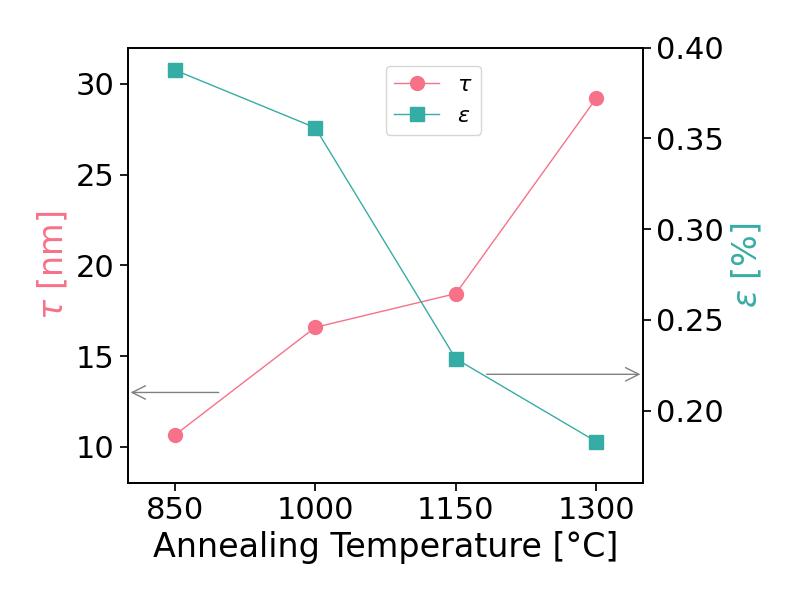}
  \caption{Crystal domain size $\tau$ and microstrain $\varepsilon$ determined from the pseudo-Voigt fit through Equations~\eqref{eq:tau} and \eqref{eq:eps}, for samples annealed at 850--1300 °C, for 1 hour.}
  \label{fig:xrd-pv}
\end{figure}

\begin{figure}[htb]
  \centering
  \includegraphics[width=0.8\linewidth]{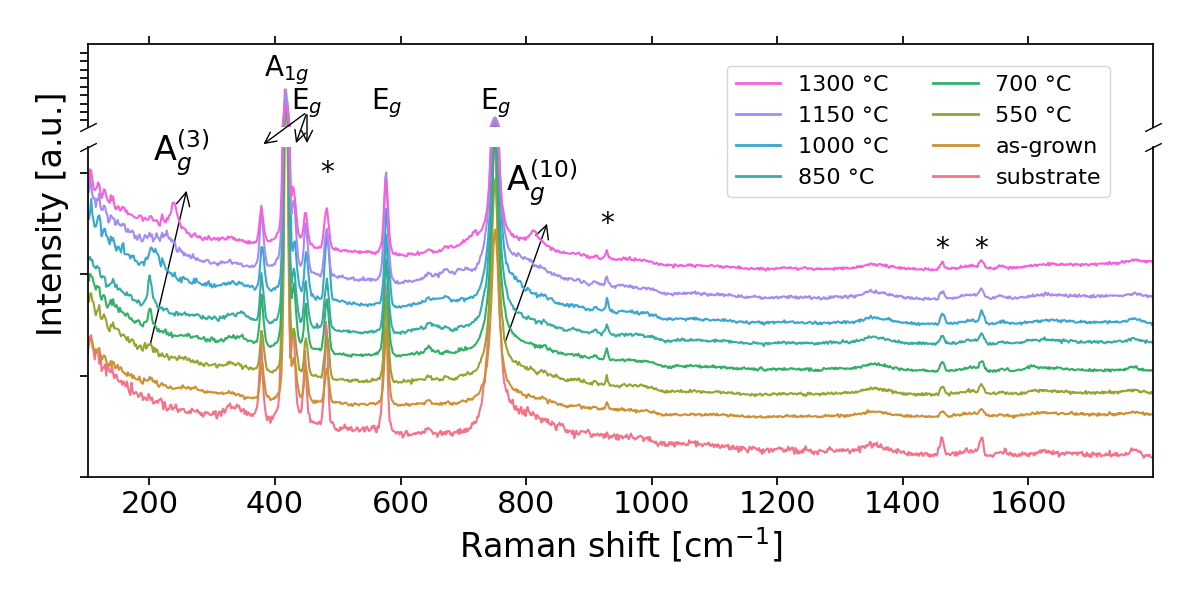}
  \caption{Raman spectra obtained for the samples annealed at 550--1300 °C, for 1 hour.}
  \label{fig:raman}
\end{figure}

\begin{figure}[htb]
  \centering
  \subfloat[Tauc plots.\label{fig:temperature-tauc}] {\includegraphics[width=0.4\linewidth]{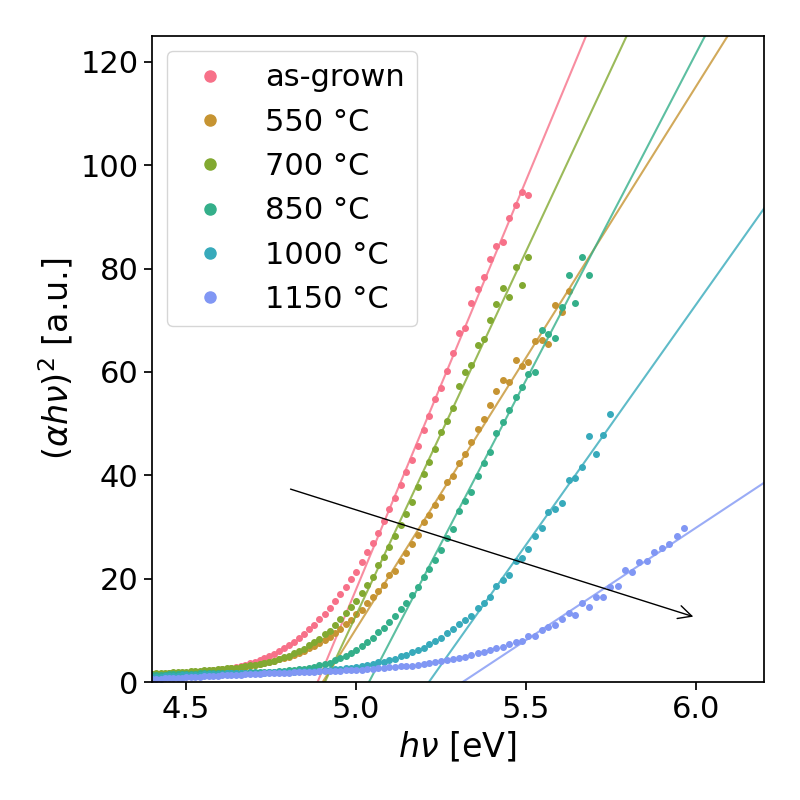}}\hfill
  \subfloat[Optical bandgaps as a function of the annealing temperature.\label{fig:41-bandgap}] {\includegraphics[width=0.6\linewidth]{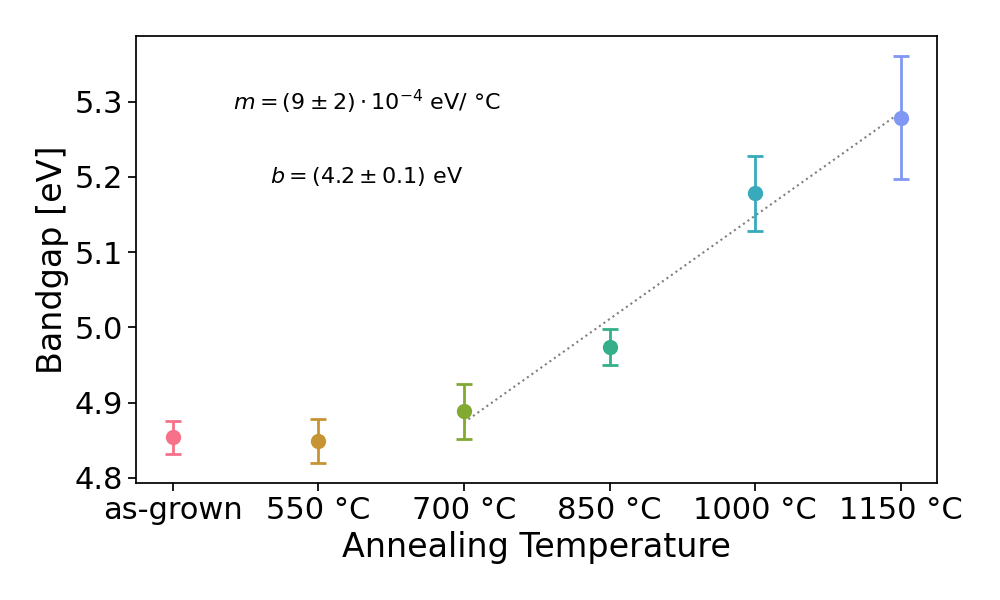}}\\
  \caption{Optical transmission analysis for as-grown and samples annealed at 550 -- 1150 °C, for 1 hour. Note that it was not possible to extract information from the sample annealed at 1300 °C due to the low intensity of the lamp at the high energies necessary to excite the wide bandgap \algao compound.} \label{fig:transmission}
\end{figure}

\begin{figure}[h]
  \centering
    \includegraphics[width=0.64\linewidth]{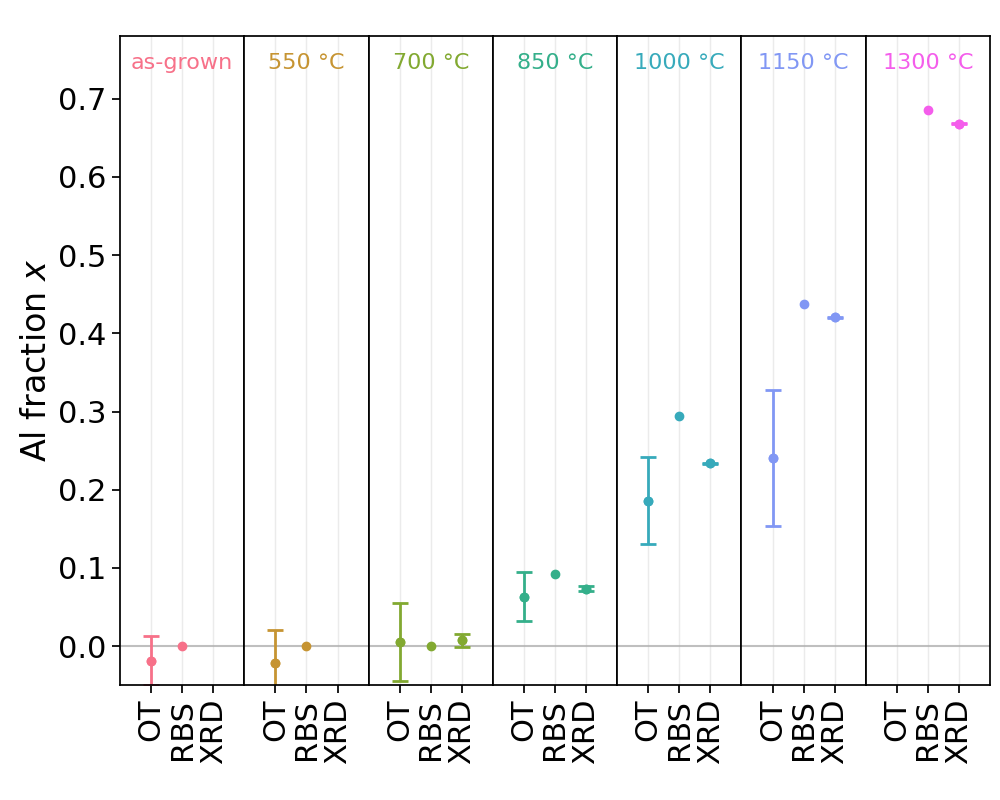}
  \caption{Comparison between the Al content of the films as determined through OT, RBS and XRD.} \label{fig:comparison}
\end{figure}
\end{widetext}

\end{document}